\documentclass{ws-ijmpcs}

\begin{document}

\markboth{KEITH SLEVIN, TOMI OHTSUKI}
{FINITE SIZE SCALING OF THE CHALKER-CODDINGTON MODEL}

%
\catchline{}{}{}{}{}
%

\title{FINITE SIZE SCALING OF THE CHALKER-CODDINGTON MODEL}

\author{KEITH SLEVIN}

\address{Department of Physics, Graduate School of Science, Osaka University\\
Machikaneyama 1-1, Toyonaka, Osaka 560-0043, Japan}

\author{TOMI OHTSUKI}

\address{Department of Physics, Sophia University\\
Kioi-cho 7-1, Chiyoda-ku, Tokyo 102-8554, Japan}

\maketitle

\begin{history}
\end{history}

\begin{abstract}
In Ref.~\refcite{slevin09}, we reported an estimate of the critical exponent for
the divergence of the localization length at the quantum Hall transition that is significantly larger
than those reported in the previous published work of other authors.
In this paper, we update our finite size scaling analysis of the Chalker-Coddington model
and suggest the origin of the previous underestimate by other authors.
We also compare our results with the predictions of L\"{u}tken and Ross.\cite{lutken07}

\end{abstract}

\ccode{PACS numbers: 73.43.$\pm$f, 71.30.+h}

\section{Introduction}	

\noindent When a strong magnetic field is applied perpendicular to an ideal two dimensional electron gas the kinetic energy of the electrons is quantized according to the formula $E_{n} =(n+1/2)\hbar \omega$.
Here, \textit{n} is a non-negative integer and $\omega$ is the cyclotron frequency
$\omega =eB/m$.
These Landau levels are highly degenerate and the density of states becomes a series of equally spaced delta functions. This degeneracy is broken by disorder and the Landau levels are broadened into Landau bands. Most of the electron states are Anderson localized with the exception of the states at the center of the Landau level where the localization length $\xi$ has a power law divergence described by a critical exponent $\nu$
\begin{equation}
\xi \sim \left|E-E_{c} \right|^{-\nu } \;.
\end{equation}
When the Fermi level is in a region of localized states, the Hall conductance is quantized in integer multiples of $e^2/h$.
This effect is known as the quantum Hall effect.\cite{klitzing80,yoshioka02} Transitions between consecutive quantized values occur when the Fermi level passes through the center of a Landau band.  This is a quantum phase transition.
It is characterized by a two critical exponents.  One is the critical exponent $\nu$ mentioned above and the other is the dynamic exponent $z$, which describes temperature dependence.

The quantum Hall transition has been the subject of careful experimental study.
The inverse of the product of the critical and dynamic exponents, $\kappa=1/\nu z$, has been measured very precisely (see Table \ref{table1}).
The value of this product appears to be quite universal; the same value has been obtained in measurements in an Al$_{x}$Ga$_{1-x}$As heterostructure\cite{li09} and in graphene.\cite{giesbers09}
The main problem is that an independent measurement of the dynamic exponent is needed to disentangle the values of the two
exponents and, unfortunately, this has been measured much less precisely.

The quantum Hall transition has also been the subject of numerous numerical studies in models of
non-interacting electrons.
In earlier work, a consensus was reached (see Table \ref{table2}) that $\nu \approx 2.4$ in
apparent agreement with experiment.
However, in 2009, we published\cite{slevin09} a numerical analysis of the Chalker-Coddington model in which we found a value of the exponent
that was about 10\% larger;
a result which has since been confirmed by other authors (see Table \ref{table3}). There now seems to be a consensus that
the previous numerical work underestimated the exponent and the apparent agreement with experiment
was a coincidence of errors.
Below we discuss the reason for the previous underestimate of the exponent.

\begin{table}[th]
\tbl{Experimental values of critical exponents for the quantum Hall transition.}
{\begin{tabular}{@{}lccc@{}} \toprule
 & $\kappa=1/z\nu$ & $\nu$ & $z$ \\ \colrule
Experiment of Li et al.\cite{li09} (Al$_{x}$Ga$_{1-x}$As) & $0.42\pm.01$ & $\approx 2.38$ & $\approx 1$ \\
Experiment of Giesbers et al.\cite{giesbers09} (graphene) & $0.41\pm.04$ & - & - \\ \botrule
\end{tabular} \label{table1}}
\end{table}

\begin{table}[th]
\tbl{Earlier estimates of the critical exponent $\nu$.}
{\begin{tabular}{@{}llll@{}} \toprule
Chalker and Coddington\cite{chalker88}      & $2.5\pm.5$        & Huckestein and Kramer\cite{huckestein90b}     & $2.34\pm.04$  \\
Mieck\cite{mieck90}                         & $2.3\pm.08$       & Huckestein\cite{huckestein92}                 & $ 2.35\pm.03$ \\
Huo and Bhatt\cite{huo92}                   & $ 2.4\pm.1$       & Lee and Wang\cite{lee96}                      & $ 2.33\pm.03$ \\
Cain et al.\cite{cain03}                    & $ 2.37\pm.02$     &                                               &        \\ \botrule
\end{tabular} \label{table2}}
\end{table}

\begin{table}[th]
\tbl{Recent estimates of the critical exponent $\nu$.}
{\begin{tabular}{@{}llll@{}} \toprule
Slevin and Ohtsuki\cite{slevin09}          & $2.593 [2.587,2.598]$      & Obuse et al.\cite{obuse10}                 & $2.55\pm .01$ \\
Dahlhaus et al.\cite{dahlhaus11} & $ 2.576\pm.03$   &  Amado et al.\cite{amado11} & $ 2.616 \pm .014$  \\ \botrule
\end{tabular} \label{table3}}
\end{table}

Another issue concerns the value of the dynamic exponent. For models of non-interacting electrons the
dynamic exponent is known exactly, $z=2$.
However, this value cannot be compared directly with the experiment.
Burmistrov et al.\cite{burmistrov11} have emphasized the distinction between the different dynamical exponents
that occur in the problem.
In the experiment of Li et al.\cite{li09} it seems clear that the dynamic exponent that is being measured describes the divergence of the phase coherence length on approaching zero temperature
\begin{equation}
    \ell_{\varphi} \sim T^{-z} \;.
\end{equation}
It also seems reasonably safe to suppose that electron-electron interactions are the source of the electron dephasing. This does not mean however that electron-electron interaction are relevant in the renormalization group (RG) sense and that the quantum Hall transition is described by a fixed point in a theory of interacting electrons.
A clear discussion of this can be found in Ref.~\refcite{burmistrov11}.
Also, as described in Ref.~\refcite{pruisken10}, it is thought that short range interactions are irrelevant in the RG sense and that only long range Coulomb interaction are relevant and would drive the system to a different interacting fixed point.

It has been pointed out to us by Alexei Tsvelik, that the value of the exponent we have found for the
Chalker-Coddington model is very close to that predicted by L\"utken and Ross.
In a series of papers (Ref.~\refcite{lutken07} and references therein) these authors have argued that modular symmetry strongly  constrains the possible critical theories of the quantum Hall transition.
While we cannot claim to understand the details of the theory of L\"utken and Ross, we attempt below to compare some of their key predictions with the results of our finite size scaling analysis.

\section{Method}
We calculated the Lyapunov exponents of the product of the transfer
matrices for the Chalker-Coddington model.\cite{chalker88}
This model describes electron localization in a two dimensional electron gas subject
to a very strong perpendicular magnetic field.
The basic assumption is that the random potential is smooth on
the scale of the magnetic length.
The electron wavefunctions are then concentrated on equipotentials of the random potential with tunneling between
equipotentials at saddle points of the potential.
In the Chalker-Coddington model this system is modeled by a network of nodes and links.
A parameter $x$, which is essentially the energy of the electrons measured in units of the Landau band
width relative
to the center of the Landau band, fixes the tunneling probability at the nodes.
A random phase distributed uniformly on $[0,2\pi)$ is attached to each link to reflect the random length of the contours of the potential.
For further details we refer the reader to the original article of Chalker and Coddington\cite{chalker88}
and to the more recent review by Kramer et al.\cite{kramer05}

\begin{figure}[th]
\centerline{\psfig{file=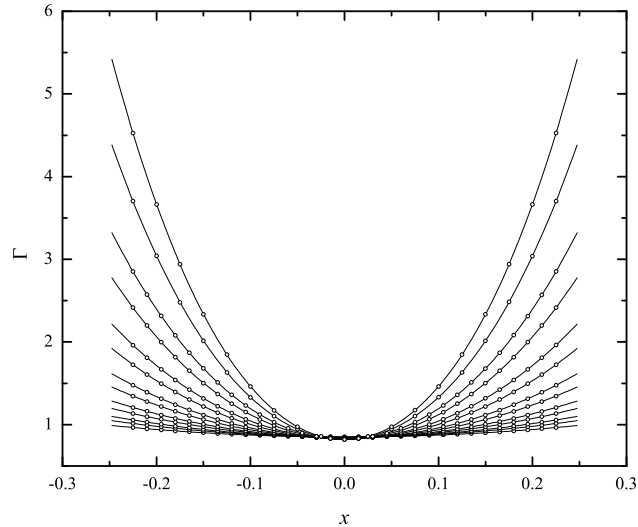,width=11cm}}
\vspace*{8pt}
\caption{The product $\Gamma$ of the smallest positive Lyapnuov exponent $\gamma$ and the number of nodes in the transverse
  direction $N=4,6,8,12,16,24,32,48,64,96,128,192,256$ for the Chalker-Coddington model.
  The error in the data is much smaller than the symbol size.
  The lines are even order polynomial fits to the data for each $N$. \label{figure1}}
\end{figure}

We considered the transfer matrix product associated with a quasi-one dimensional geometry with $N$ nodes in the transverse direction
and $L$ nodes in the longitudinal direction.\footnote{For the detailed formulae see Ref.~\refcite{slevin09}.}
The Lyapunov exponents of this random matrix product were estimated using the standard method.\cite{shimada79,mackinnon83}
The Lyapunov exponents are defined by taking the limit $L\rightarrow \infty$. By truncating the matrix product at a finite
$L$, an estimate of the Lyapunov exponents was obtained.
The sample to sample fluctuations of this estimate decrease with the inverse of the square root of $L$.
We performed a single simulation for each pair of $x$ and $N$ and truncated the transfer matrix product at a value of $L$ that
allowed estimation of the smallest positive Lyapunov exponent $\gamma$ with a precision of $0.03\%$,
except for the largest values of $N=192$ and $256$ where the precision was relaxed to either $0.05\%$ or $0.1\%$.
To ensure that simulations for different pairs of $x$ and $N$ were independent,
we used the Mersenne Twister pseudo-random number generator MT2203 of Matsumoto et al.\cite{matsumoto99} provided in the Intel Math Kernel Library.
All the simulations used a common seed.
Independence was ensured by the use of a unique stream number for each simulation.
We imposed periodic boundary conditions in the transverse direction for which choice
the Lyapunov exponents are even functions of $x$.
It is known that there is a critical point at the center of the Landau band, $x=0$, and that, when the
Fermi energy is driven through this point, the transition between Hall plateaux occurs.

\begin{figure}[th]
\centerline{\psfig{file=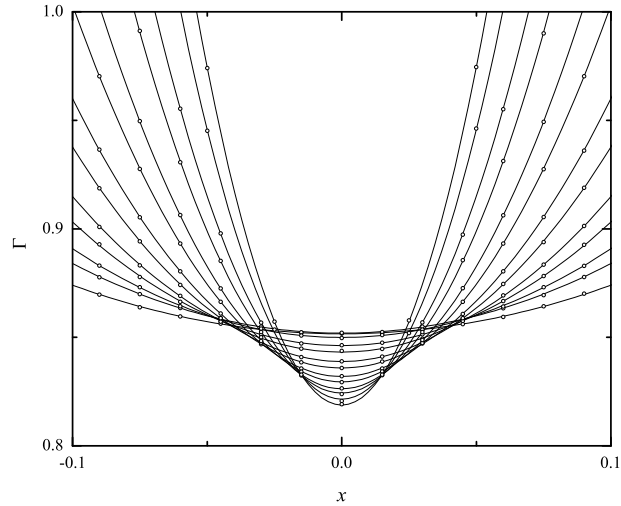,width=10.5cm}}
\vspace*{8pt}
\caption{The same data as in Fig.~\ref{figure1} but focussing on the critical point at $x=0$.
The residual variation of $\Gamma$ with $N$ at $x=0$ is due to irrelevant scaling variables. \label{figure2}}
\end{figure}

To extract estimates of critical exponent and other quantities we used finite size scaling.\footnote{This method was first applied to Anderson localization at about the same time by Pichard and Sarma\cite{pichard81a,pichard81b} and by MacKinnon and Kramer\cite{mackinnon83,mackinnon81}. See Ref.~\refcite{amit05} for a pedagogical discussion.}
In this method, the behavior of the dimensionless quantity
\begin{equation}
    \Gamma \left(x,N \right) = \gamma N \;,
\end{equation}
is analyzed as a function of both $x$ and $N$.
In the absence of any corrections to scaling we would expect this behavior to be described by
the following finite size scaling law
\begin{equation}\label{eq:FSSnocorrections}
    \Gamma = F\left( N^{2 \alpha} x^2 \right) \;,
\end{equation}
where $F$ is an a priori unknown but universal scaling function and
\begin{equation}
    \alpha = 1/\nu \;.
\end{equation}
Note that we have imposed the condition that $\Gamma$ must be an even function of $x$.

The actual behavior of $\Gamma$ as a function of $x$ for different $N$ is shown in Fig.~\ref{figure1}
and, in more detail around $x=0$, in Fig.~\ref{figure2}.
(The lines in the figures are polynomial fits. They will be discussed below.)
According to Eq.~(\ref{eq:FSSnocorrections}) curves for different $N$ should have a common crossing point
at $x=0$.
However, it is clear from Fig.~\ref{figure2} that this is only approximately correct and that
$\Gamma$ is not exactly independent of $N$ at $x=0$ but varies by a several percent over the range of $N$ studied.
These corrections to scaling arise because of the presence of irrelevant scaling variables. These are variables
with negative scaling exponents. Their effect is negligible for large $N$ but their presence may lead to significant
corrections at small $N$. This is consistent with what we see in Fig.~\ref{figure2}.

To take account of corrections to scaling, we follow Ref.~\refcite{huckestein94b} and Ref.~\refcite{slevin99a} and
generalize the finite size scaling law
\begin{equation}\label{eq:FSSwithcorrections}
    \Gamma = F\left( N^{2 \alpha} v_0( x ), N^{y_1} v_1( x ), N^{y_2} v_2( x ), \cdots \right) \;.
\end{equation}
In Eq. (\ref{eq:FSSwithcorrections}), $v_0$ is the relevant scaling variable and $v_1,v_2,\cdots$ are the
irrelevant scaling variables. The associated exponents $y_1, y_2, \cdots$ are negative.
The inclusion of irrelevant corrections permits the residual $N$ dependence at $x=0$ seen in Fig.~\ref{figure2} to be modeled.
This form also allows for additional corrections due to non-linearities of the scaling variables as functions of $x$.
To impose the condition that $\Gamma$ must be an even function of $x$ we restrict all the scaling variables to be even
functions of $x$.
In addition, since the critical point is at $x=0$, we impose the condition that $v_0( 0  ) = 0$.
To fit the data, the function $F$ is expanded as a Taylor series in all its arguments, and similarly the scaling
variables.
The coefficients in the Taylor series, together with the various exponents, play the role of fitting parameters.\footnote{Some extra conditions on the coefficients must be imposed to ensure that the model to be fitted to the data is unambiguous.}
The orders of truncation of the Taylor series are chosen sufficiently large to obtain an acceptable fit of the data (as measured using the $\chi^2$-statistic and the goodness of fit probability).
We have attempted this procedure with both one and two irrelevant corrections.
Unfortunately, a stable fit of the data has eluded us.
We have found that several fits of the data are possible.
However, these fits do not yield mutually consistent estimates of the critical exponent.

\section{Rudimentary finite size scaling}

\begin{figure}[t]
\centerline{\psfig{file=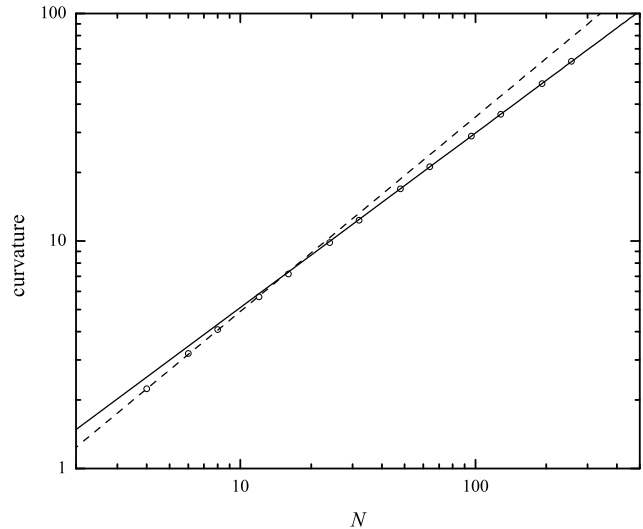,width=11cm}}
\vspace*{8pt}
\caption{The curvature $C$ at $x=0$ (see Eq. (\ref{eq:curvature})), obtained from the polynomial fits shown in Figure \ref{figure1}, plotted as a function of $N$. The solid line is a straight line
  fit to the data for the largest five values of $N$. The slope corresponds to $\nu=2.607$. The dashed line is a straight line with slope corresponding to $\nu=2.34$ and passing though the $N=4$ data point.
  The error in the data for the curvature is much smaller than the symbol size.
  \label{figure3}}
\end{figure}

To circumvent the difficulties described in the previous section we resorted to a less sophisticated approach
in which we abandoned the attempt to fit all the data in a single step.
Instead, we fitted the data for each $N$ independently to an even polynomial of $x$.
For each $N$ the order of the polynomial was chosen just large enough to give an acceptable goodness of fit.
For $N=4$, a quadratic was sufficient, while for $N=256$ a sixth order polynomial was required.
From these polynomials we estimated the curvature $C$ of $\Gamma$ at $x=0$.
The results are plotted in Fig.~\ref{figure3}.
The precision of the estimation of the curvature varies between $0.07\%$ and $0.28\%$.
According to (\ref{eq:FSSnocorrections}) the curvature at $x=0$ should vary with $N$ as a power law,
\begin{equation}\label{eq:curvature}
    C \equiv \left. \frac{d^2 \Gamma }{d x^2}\right|_{x=0} \propto N^{2\alpha} \;.
\end{equation}
This, of course, neglects corrections to scaling due to irrelevant variables and, indeed,
a straight line fit to \emph{all} the data does not yield an acceptable goodness of fit.
However, if data for $N\leq 48$ are excluded, acceptable goodness of fits are obtained.
The estimates of the critical exponent obtained in this way are tabulated in Table~\ref{table4}.

In Fig.~\ref{figure3} we have also plotted two lines. One is a solid line that corresponds to the straight line
fit for $64\leq N \leq 256$.
The second is a dashed line with a slope corresponding
to the estimate of Huckestein\cite{huckestein94b} of the critical exponent and passing through the datum for the curvature at $N=4$.
While the main effect of the irrelevant corrections is the $N$ dependent shift in
the ordinate that is clearly visible in Fig.\ref{figure2}, a smaller but not negligible effect on
the curvature is also apparent in Fig.\ref{figure3}.
In our opinion, this is the reason why the critical exponent was underestimated in previous work.
The precision of the numerical data was insufficient, and the range of $N$ considered too small, for the
irrelevant correction to be properly taken into account.

\begin{table}[th]
\tbl{Estimates of the critical exponent $\nu$ obtained from linear fits of $\ln C$ versus $\ln N$.}
{\begin{tabular}{@{}lcc@{}} \toprule
range of $N$ considered                     & $\nu$                 & Confidence intervals ($\pm 2 \sigma$) \\ \colrule
$64 \leq N \leq 256$                        & $2.607$               & $[2.598,2.615]$         \\
$96 \leq N \leq 256$                        & $2.599$               & $[2.583,2.616]$         \\
$128 \leq N \leq 256$                       & $2.590$               & $[2.586,2.614]$         \\ \botrule
\end{tabular} \label{table4}}
\end{table}

\section{Comparison with the predictions of L\"{u}tken and Ross}
The predictions of the theory of L\"{u}tken and Ross that can be compared with the present work are for the critical exponent and
the leading irrelevant exponent.
Their theory contains a single unknown parameter, the central charge $c$. In terms of this parameter they predict that
the critical exponent is\footnote{We noticed an error (a factor of 2) in Ref.~\refcite{lutken07}.
We thank Graham Ross for confirming this.}
\begin{equation}\label{eq:LRexponent}
    \nu = \frac{10.42050633345819\cdots}{c} \;.
\end{equation}
In addition they predict that the leading irrelevant exponent and the critical exponent are related by
\begin{equation}\label{eq:LRirrelevant}
    y = -\alpha = -1 /\nu \;.
\end{equation}
While we are not aware of any justification for this, assuming $c=4$
 gives
\begin{equation}
    \nu = 2.60512\cdots
\end{equation}
The most precise estimate in Table~\ref{table4} is
\begin{equation}
    \nu = 2.607 \pm 0.004 \;.
\end{equation}
In fact, all of the estimates in Table~\ref{table4} are consistent with the L\"{u}tken and Ross value. Turning to the
irrelevant exponent, the situation is, unfortunately, much less clear.
In Table~\ref{table5} we tabulate some previous estimates of the irrelevant exponent.
The L\"{u}tken and Ross prediction is
\begin{equation}\label{eq:LRy}
    y= -0.383859\cdots
\end{equation}
The estimate of Huckestein\cite{huckestein94b} is consistent with this but subsequent estimates by Wang et al.\cite{wang98}
are somewhat ambiguous. In our opinion, further confirmation is needed before reaching a conclusion.

\begin{table}[th]
\tbl{Estimates from the literature of the leading irrelevant exponent $y$ and the quantities
used to estimate it.}
{\begin{tabular}{@{}lll@{}} \toprule
Huckestein \cite{huckestein94b}  & $y = -0.38\pm .04$    &   Lyapunov exponents    \\
Wang et al.\cite{wang98}               & $y\approx -0.52$      & geometric average of the two-terminal conductance    \\
Wang et al.\cite{wang98}               & $y\approx -0.72$      & arithmetic average of the two-terminal conductance    \\ \botrule
\end{tabular} \label{table5}}
\end{table}

\begin{table}[th]
\tbl{Fits to the 379 data points shown in Figure~\ref{figure1}.}
{\begin{tabular}{@{}cccccccc@{}} \toprule
$n_0$   & $n_1$   & number of parameters     & $\chi^2$  & goodness of fit & $\Gamma_c$      \\  \colrule
 3      & 3       & 16        & 398       & 0.10      & $0.807 \pm .0005$          \\
 3      & 4       & 20        & 347       & 0.66      & $0.804 \pm .0015$          \\
 3      & 5       & 24        & 343       & 0.67      & $0.801 \pm .0035$          \\  \botrule
\end{tabular} \label{table6}}
\end{table}

\begin{table}[th]
\tbl{Estimates of $\Gamma_c$ obtained using published estimates of $\alpha_0$ and Eq.~(\ref{eq:confinv})}
{\begin{tabular}{@{}lcc@{}} \toprule
                                & $\alpha_0$                & $\Gamma_c$                \\  \colrule
Obuse et al.\cite{obuse08}      & $2.2617 \pm 0.0006$       & $0.8222 \pm .0019$        \\
Evers et al.\cite{evers08}      & $2.2596 \pm 0.0004$       & $0.8156 \pm .0013 $       \\  \botrule
\end{tabular} \label{table7}}
\end{table}

As mentioned above an unambiguous fit using Eq. (\ref{eq:FSSwithcorrections}) has not proved possible. The
best we can do at present is to check the consistency of the L\"{u}tken and Ross values with our data by fixing both $\nu$ and $y$ to these values when fitting.
A series of such fits are tabulated in Table \ref{table6}. In the fits only a single irrelevant variable is assumed and
non-linearities in the scaling variables are ignored, i.e.
\begin{equation}
    v_0 = v_{02} x^2 \; \;, \; \; v_1 = v_{10} \;.
\end{equation}
The scaling function is expanded as a Taylor series to order $n_0$ in the relevant field and order $n_1$ in the
irrelevant field.
One of the important quantities that can be estimated in this way is $\Gamma_c$, which is defined by
\begin{equation}
    \Gamma_c = \lim_{N\rightarrow\infty} \Gamma \left( x=0,N \right) = F \left( 0,0,\ldots \right) \;.
\end{equation}
The quantity is significant because, if the quantum Hall critical theory has conformal symmetry, it is related to the
multi-fractal exponent $\alpha_0$ that occurs in the multi-fractal analysis of the wavefunction distribution at the critical point
by
\begin{equation}\label{eq:confinv}
    \Gamma_c = \pi \left( \alpha_0 - 2 \right) \;.
\end{equation}
Some estimates of $\Gamma_c$ obtained using published estimates of $\alpha_0$ and Eq.~(\ref{eq:confinv}) are tabulated in Table~\ref{table7}.
Our numerical estimate of $\Gamma_c$ is not completely consistent with those in the table.
The reason for this is not yet clear.

\section{Discussion}
The agreement between our estimate for the critical exponent and prediction of L\"{u}tken and Ross is tantalizing but is it accidental, or does it have a deeper significance?
A more precise numerical estimate of the irrelevant exponent is clearly highly desirable.
In addition, the prediction of  L\"{u}tken and Ross for the flow diagram in the $\left(\sigma_{xy},\sigma_{xx} \right)$
plane should also be amenable to numerical verification.
We also need to know if a central charge $c=4$ is physically justified.

Quite apart from whether or not the L\"{u}tken and Ross theory is exact for non-interacting electrons, the important
question remains of clarifying the role of the electron-electron interactions in the observed critical behavior at the
quantum Hall transition.
More work along the lines of Ref.~\refcite{huckestein99} might be very helpful in this regard.

\section*{Acknowledgments}
We would like to thank Alexei Tsvelik for bringing the work of L\"{u}tken and Ross to our attention.
This work was supported by Grant-in-Aid 23540376 and Korean WCU program Project No. R31-2008-000-10059-0.

\bibliographystyle{prsty2}
\bibliography{references}

\begin{thebibliography}{10}

\bibitem{slevin09}
K. Slevin and T. Ohtsuki, Physical Review B {\bf 80},  041304(R)  (2009).

\bibitem{lutken07}
C.~A. Lutken and G.~G. Ross, Physics Letters B {\bf 653},  363  (2007).

\bibitem{klitzing80}
K.~v. Klitzing, G. Dorda, and M. Pepper, Physical Review Letters {\bf 45},  494
   (1980).

\bibitem{yoshioka02}
D. Yoshioka, {\em The quantum Hall effect}, {\em Springer series in solid-state
  sciences,} (Springer, Berlin ; New York, 2002).

\bibitem{li09}
W. Li {\it et~al.}, Physical Review Letters {\bf 102},  216801  (2009).

\bibitem{giesbers09}
A.~J.~M. Giesbers {\it et~al.}, Physical Review B (Condensed Matter and
  Materials Physics) {\bf 80},  241411  (2009).

\bibitem{chalker88}
J.~T. Chalker and P.~D. Coddington, Journal of Physics C: Solid State Physics
  {\bf 21},  2665  (1988).

\bibitem{huckestein90b}
B. Huckestein and B. Kramer, Physical Review Letters {\bf 64},  1437  (1990).

\bibitem{mieck90}
B. Mieck, EPL (Europhysics Letters) {\bf 13},  453  (1990).

\bibitem{huckestein92}
B. Huckestein, EPL (Europhysics Letters) {\bf 20},  451  (1992).

\bibitem{huo92}
Y. Huo and R.~N. Bhatt, Physical Review Letters {\bf 68},  1375  (1992).

\bibitem{lee96}
D.-H. Lee and Z. Wang, Philosophical Magazine Letters {\bf 73},  145   (1996).

\bibitem{cain03}
P. Cain, R.~A. Romer, and M.~E. Raikh, Physical Review B {\bf 67},  075307
  (2003).

\bibitem{obuse10}
H. Obuse {\it et~al.}, Physical Review B {\bf 82},  035309  (2010).

\bibitem{dahlhaus11}
J.~P. Dahlhaus, J.~M. Edge, J. Tworzydlo, and C.~W.~J. Beenakker, Physical
  Review B {\bf 84},  115133  (2011).

\bibitem{amado11}
M. Amado {\it et~al.}, Physical Review Letters {\bf 107},  066402  (2011).

\bibitem{burmistrov11}
I.~S. Burmistrov {\it et~al.}, Annals of Physics {\bf 326},  1457  (2011).

\bibitem{pruisken10}
A.~M.~M. Pruisken, International Journal of Modern Physics B {\bf 24},  1895
  (2010).

\bibitem{kramer05}
B. Kramer, T. Ohtsuki, and S. Kettemann, Physics Reports {\bf 417},  211
  (2005).

\bibitem{shimada79}
I. Shimada and T. Nagashima, Progress of Theoretical Physics {\bf 61},  1605
  (1979).

\bibitem{mackinnon83}
A. MacKinnon and B. Kramer, Zeitschrift f\"{u}r Physik B Condensed Matter {\bf
  53},  1  (1983).

\bibitem{matsumoto99}
M. Matsumoto and T. Nishimura,  in {\em Monte Carlo and Quasi-Monte Carlo
  Methods 1998}, edited by H. Niederreiter and J. Spanier (Springer, Berlin ;
  New York, 1999), pp.\ 56--69.

\bibitem{pichard81a}
J.~L. Pichard and G. Sarma, Journal of Physics C: Solid State Physics  L127
  (1981).

\bibitem{pichard81b}
J.~L. Pichard and G. Sarma, Journal of Physics C: Solid State Physics  L617
  (1981).

\bibitem{mackinnon81}
A. MacKinnon and B. Kramer, Physical Review Letters {\bf 47},  1546  (1981).

\bibitem{amit05}
D.~J. Amit and V. Martin-Mayor, {\em Field theory, the renormalization group,
  and critical phenomena}, 3rd ed. ed. (World Scientific, New Jersey ; London,
  2005).

\bibitem{huckestein94b}
B. Huckestein, Physical Review Letters {\bf 72},  1080  (1994).

\bibitem{slevin99a}
K. Slevin and T. Ohtsuki, Physical Review Letters {\bf 82},  382  (1999).

\bibitem{wang98}
X. Wang, Q. Li, and C.~M. Soukoulis, Physical Review B {\bf 58},  3576  (1998).

\bibitem{obuse08}
H. Obuse {\it et~al.}, Physical Review Letters {\bf 101},  116802  (2008).

\bibitem{evers08}
F. Evers, A. Mildenberger, and A.~D. Mirlin, Physical Review Letters {\bf 101},
   116803  (2008).

\bibitem{huckestein99}
B. Huckestein and M. Backhaus, Physical Review Letters {\bf 82},  5100  (1999).

\end{thebibliography}

\end{document}